\documentstyle[psfig,epsfig]{mn}

\def\Lsun{\hbox{$\rm\thinspace L_{\odot}$}}
\def\Msun{\hbox{$\rm\thinspace M_{\odot}$}}
\def\yr{{\rm\thinspace yr}}
\def\Mpc{{\rm\thinspace Mpc}}
\def\kmps{\hbox{$\km\s^{-1}\,$}}
\def\cm{{\rm\thinspace cm}}
\def\g{{\rm\thinspace g}}
\def\s{{\rm\thinspace s}}
\def\km{{\rm\thinspace km}}
\def\Msunpyr{\hbox{$\Msun\yr^{-1}\,$}}
\def\kmpspMpc{\hbox{$\kmps\Mpc^{-1}$}}
\def\Gyr{{\rm\thinspace Gyr}}
\newcommand{\apj}{ApJ, }
\newcommand{\ApJ}{ApJ, }
\newcommand{\apjs}{ApJ Suppl., }
\newcommand{\aj}{AJ, }

\newcommand{\mn}{MNRAS, }
\newcommand{\mnras}{MNRAS, }

\def\T_d{$T_d = 40$ K\phantom{a}}

\def\obs{38\phantom{a}}
\def\det{8\phantom{a}}

\def\stat{35\phantom{a}}
\def\less_10{19\phantom{a}}
\def\great_10{8\phantom{a}}
\def\great_12{3\phantom{a}}
\def\total_great_10{11\phantom{a}}
\def\omont_det_rate{$\sim 30\%$\phantom{a}}
\def\jcmt_det_rate{$\sim 24\%$\phantom{a}}
\def\sn_peak{$\sim 0.3$}

\title{ The SCUBA Bright Quasar Survey (SBQS): 
850micron observations of the z$\ga$4 sample}
\author[K.G. Isaak, Robert S. Priddey, Richard G. McMahon et al.]
{Kate G. Isaak$^{1}$, Robert S. Priddey$^{2,3}$, Richard G. McMahon$^{2}$, 
Alain Omont$^{4}$,
\and Celine Peroux$^{2}$, Robert G. Sharp$^{2}$ 
and Stafford Withington$^1$\\
$^{1}$ Cavendish Astrophysics, University of Cambridge, Cambridge CB3 0HE; 
$^{2}$ Institute of Astronomy, University of Cambridge, \\
Cambridge CB3 OHF; $^{3}$ Blackett Laboratory, Imperial College, 
London SW7 2BW; $^{4}$ Institute of Astrophysics, Paris\\
email: isaak@mrao.cam.ac.uk[KGI]; r.priddey@ic.ac.uk[RSP];
rgm@ast.cam.ac.uk[RGM]; omont@iap.fr[AO]; celine@ast.cam.ac.uk[CP];\\
rgs@ast.cam.ac.uk[RGS];stafford@mrao.cam.ac.uk[SW]}

\date{submitted April, 2001
      accepted 7th September, 2001}

\pagerange{\pageref{firstpage}--\pageref{lastpage}}
\pubyear{2001}

\begin{document}
\maketitle

\label{firstpage}

\begin{abstract}     
We present initial results of a new, systematic search 
for massive star-formation in the host galaxies of the most luminous and
probably most massive z$\ga$4 radio-quiet quasars
($M_B \leq -27.5$; $\nu L_{\nu}(\rm 1450\AA)>10^{13}$\Lsun).
The survey, undertaken at $850\mu m$ using SCUBA at the 
James Clerk Maxwell Telescope (JCMT), has a target sensitivity limit of
$3\sigma\sim$10mJy, set to identify sources suitable for detailed 
follow-up e.g. continuum mapping and molecular line diagnostics. 
A total of \obs z$\ga$4 radio-quiet quasars have been 
observed at 850$\mu m$, of which  
\det were detected ($>$3$\sigma$) with $S_{850\mu m}\ga 10$mJy
(submillimetre-loud). 
The new detections almost triple the number of optically 
selected, submillimetre-loud z$\ga$4 radio-quiet quasars known to date.
We include a detailed description of how our quasar sample is defined
in terms of radio and optical properties. 
As a by-product of our
selection procedure we have identified 17 radio-loud quasars with
z$\ga$4. 

There is no strong evidence for trends in either detectability or 
850$\mu m$ flux with absolute magnitude, $M_B$.
We find that the 
weighted mean flux of the undetected sources is $2.0 \pm 0.6$mJy, 
consistent with an earlier estimate of $\sim3$mJy 
based on more sensitive observations of a sample 
z$\ga$4 radio-quiet quasars (McMahon {\it et al.},1999). This corresponds
to an inferred starformation rate of $\sim$1000\Msunpyr,
similar to Arp220.
The typical starformation timescale for the submillimetre-bright 
sources is $\sim1\Gyr$, 10 times longer than the typical accretion-driven 
e-folding timescale of $\sim5\times10^7$ years.
Our 850$\mu m$ detection of the $z=4.4$ quasar PSS J1048$+$4407 
when analysed in conjunction with 1.2mm single-dish 
and interferometric observations 
suggests that this
source is resolved on angular scales of 1-2\arcsec (6-12 kpc).
In addition, we present a new optical spectrum of this source,
identifying it as a broad absorption line (BAL) quasar. The new redshift
is outside that covered in a recent CO line search 
by Guilloteau {\it et al.}, (1999), highlighting the need for 
accurate redshifts for the obervation and interpretation of 
high-redshift line studies. 
\end{abstract}

\begin{keywords}
high-redshift star-formation ; radio-quiet quasars; quasar host galaxies
\end{keywords}

\section{Introduction}
\begin{figure*}
\psfig{figure=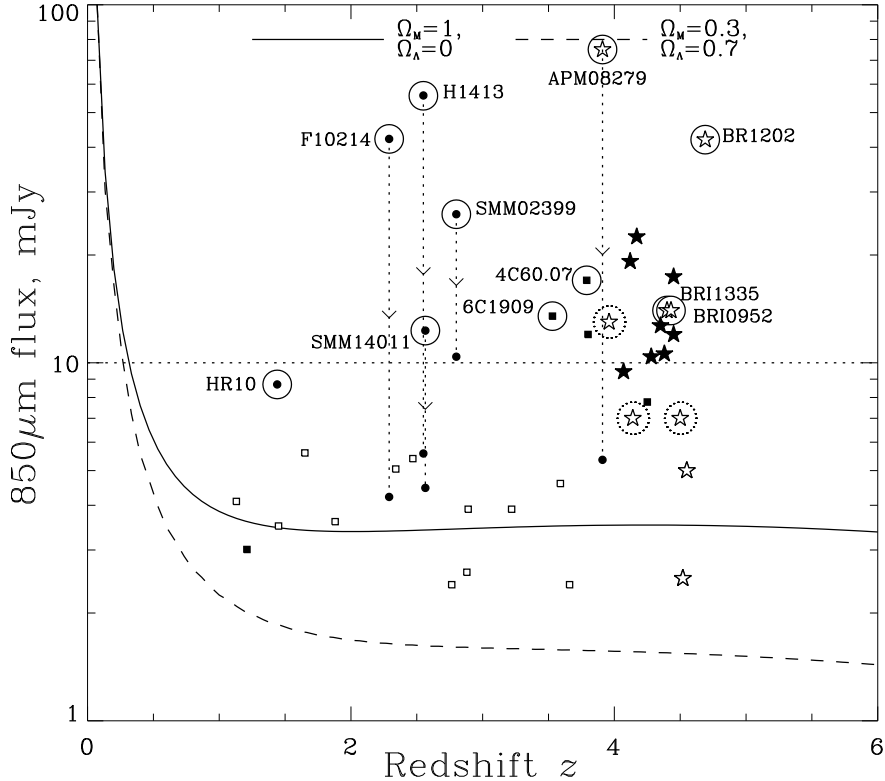,width=150mm}
\caption[]{A plot of the $850\mu m$ flux vs. redshift for  
high-z objects with observed $850\mu m$ fluxes and accurately determined
redshifts. Stars denote detections of z$\ga$4 radio-quiet quasars, with
solid stars representing detections presented in this paper, and open
stars sources taken from previous work.
Squares denote detections of radio galaxies -- unfilled taken from 
Archibald et al, 2000, and filled from the literature.
Solid circles denote a selection of other high-z objects. 
Objects for which CO observations exist are ringed: 
solid rings -- CO detections; dashed rings -- CO non-detections.
Lensing corrections are indicated by vertical dotted lines. 
The two curves trace the estimated flux that an 
Arp220-like object (based on a fit to observed FIR fluxes of Arp220) 
would have at different redshifts
(solid line: $\Omega_M=1; \Omega_\Lambda = 0.$; 
dashed line: $\Omega_M = 0.3; \Omega_\Lambda = 0.7$)
The dashed horizontal $3\sigma$ 
limit of this survey.}
\label{fig_detectability}
\end{figure*}

It is becoming increasingly clear that star-formation with its
complementary photometric and spectroscopic signatures provides
observers with a range of probes with which to study formation and
evolution of galaxies in the early Universe. Particularly important
to such studies are the millimetre-/submillimetre-/far-infrared-
wavebands, a wavelength range in which clear fingerprints of both warm
dust, heated by UV starlight from massive young stars, as well as
atomic and molecular line emission, excited in regions of on-going
starformation, can be seen.
Recent work by Wright \& Reese \shortcite{Wright00} has further
illustrated the importance of this waveband, demonstrating that roughly
half the radiation produced by starlight in galaxies is absorbed by
dust and reradiated in the far-infrared/submillimetre-waveband.
Likewise, COBE observations have shown that the cosmological
starformation rates inferred from UV- and optical studies 
(eg. Madau {\it et al.}, (1998))
are more than a factor of two lower than those required to
produce the observed far-infrared background, suggesting that there
exists a population of star-forming galaxies that are hidden at
optical wavelengths.

How can one use the (sub)millimetre-waveband to study starformation, both 
qualitatively and quantitatively, in the 
high-redshift universe?
Photometric measurements of the (sub)millimetre-wave emission
constrain the rest-frame spectral energy distribution -- with more
than one (sub)millimetre flux it is possible to evaluate a (sub)millimetre
spectral index, thus confirming or refuting a thermal origin 
of the emission (eg. Hughes {\it et al.}\shortcite{Hughes93},
Isaak {\it et al.}\shortcite{Isaak94}). With sufficient data, one can 
also determine a best-fit dust temperature, from which a 
bolometric (sub)millimetre-/far-infrared luminosity can be derived. 
Establishing the original energy source responsible for this reprocessed 
emission is considerably more difficult 
(eg. Sanders {\it et al.}, \shortcite{Sanders89}),  as the continuum 
spectral signatures of dust-obscured AGN and massive starbursts are very 
similar, indeed, with current telescopes indistinguishable at 
(sub)millimetre-wavelengths. If one assumes that some fraction of the 
(sub)millimetre/far-infrared luminosity arises from reprocessed UV-light 
from massive stars, then it is possible to infer starformation rates also.
The presence and detection of the molecular gas tracer CO, found ubiquitously
in nearby starburst galaxies, points to future starformation potential.
In addition spectral line measurements provide important kinematic and 
dynamic information. 
Whilst thermal emission from dust out to high redshift 
can now be detected with relative ease due to the new 
generation of sensitive multi-pixel millimetre \cite{Kreysa98} and 
submillimetre-wave arrays \cite{Holland99},  
detecting CO is more difficult, requiring
a very accurate measure of the source redshift because of the limited 
bandwidth ($\Delta\nu\sim1 GHz$), and thus small fractional bandwidth
at 100-200 GHz ($<1\%$),  of the current generation of millimetre-wave 
receivers. 

\section{Survey design}
\subsection{Choice of targets} 
 Recent attempts to study starformation in the high redshift Universe 
have focused on identifying target objects in two quite distinct ways.
The first of these -- deep, unbiased surveys of blank sky -- only
recently became possible with the introduction of sensitive,
multi-pixel (sub)millimetre-wave bolometer arrays.
Surveys undertaken
using SCUBA (Submillimetre Continuum Bolometer Array \cite{Holland99}) at the
James Clerk Maxwell Telescope (eg. Hughes {\it et
  al.}\shortcite{Hughes96}, Barger {\it et al.}\shortcite{Barger96}, 
Smail {\it et al.}\shortcite{Smail97},
Eales {\it et al.}\shortcite{Eales99}) have been particularly
successful, identifying some tens of submillimetre-selected sources
with $850\mu m $ fluxes $S_{850\mu m}>5$mJy. Classifying these
sources, in particular assigning redshifts, is complicated by
(i) the limited accuracy with which one can
determine the source positions using a single-dish telescope (ii)
the large number of potential, though very faint, optical counterparts
revealed in deep observations of the submillimetre-wave source fields.
As a result it has been possible to begin to derive the starformation
properties of only handful of submillimetre-selected sources (eg.
Frayer {\it et al.}\shortcite{Frayer98,Frayer00}).

In contrast, objects with known redshift provide a very convenient starting 
point from which to base searches for and studies of 
star-formation at high-redshift. The first submillimetre-wave observations of
high redshift objects were made in the early 1990s 
\cite{Clements92,Barvainis92}. 
Following on from the successful detection both of the (sub)millimetre
continuum (by Clements {\it et al.}, \shortcite{Clements92}) and 
CO emission (by Brown \& VandenBout \shortcite{Brown91} and 
subsequently Solomon, Radford and Downes \shortcite{Solomon92}) from
the ultra-luminous infrared galaxy IRAS F10214+4724 at $z\sim2.3$,  
(sub)millimetre studies of the continuum and CO line emission at high 
redshift of a small sample of radio-quiet quasars taken 
from the Cambridge APM BR/BRI $z\sim4$ QSO survey 
\cite{Irwin91,SMIH96,SIMH01}, were made. At the time, these quasars were
amongst the most distant known. 
With the combination of the $800\mu m/1100\mu m$ measured with UKT14 
\cite{Duncan90} on the JCMT \cite{Isaak94} and the $1300 \mu m$  
fluxes measured at IRAM-30m \cite{McMahon94}, it was possible to 
show that the (sub)millimetre continuum spectral 
indices were consistent with 
thermal emission. The (sub)millimetre emission 
thus indicates the presence of massive quantities of warm dust ($T_d\sim50$K)
at redshifts of $z\sim4$, 
synthesized, and possibly heated, by bursts of
star-formation in the host galaxies. 
Large quantities of molecular gas, $M_{H_2} > 10^{10} M_{\odot}$ as 
traced by CO (eg. Ohta {\it et al.}\shortcite{Ohta96},
Omont {\it et al.}\shortcite{Omont96a}, Guilloteau {\it et al.} 
\shortcite{Guilloteau99}), were also detected in the host galaxies of 
a handful of quasars at $z>4$. 
The detected gas masses are comparable to the total 
dynamical mass of the galaxy, a situation seen in galaxies 
undergoing their first bouts of star-formation (eg. Downes \& Solomon 
\shortcite{Downes98}).

\subsection{Survey definition}
The number of objects from which {\it both} dust and CO emission have been 
detected is small, and so it is currently not 
possible to differentiate between 
observed/derived properties that are representative of objects at 
high redshift, and those that are extreme. An efficient way by which to 
increase the number of massive star-forming galaxies that can be 
readily studied needs to be identified. 
Shown in Figure \ref{fig_detectability} is a plot of the {\it observed} 
$850\mu m$ flux 
vs. redshift for a selection of high-redshift objects with accurately 
determined redshifts. 
Looking at $z\ga4$: 
each source from which CO emission has been detected at greater than 5$\sigma$
(see Guilloteau {\it et al.}, 1999 and references therein) has  
$S_{850\mu m} >$ 10mJy, and conversely CO emission has been detected from 4 of the 
5 objects with $S_{850\mu m} >$ 10mJy \footnote{CO has not detected in
BR B1117$-1$1329,
which  has $S(850\mu m) = 13\pm1$ mJy, Buffey {\it et al.}, in prep.)}. 
This suggests an observationally 
imposed cutoff for successful CO detections of $S_{850\mu m}\sim10$mJy.
If we assume that by 
$z\sim5$ 
(0.9(1.2)$\Gyr$ after recombination, using an Einstein De Sitter cosmology, 
with $H_o=50$ \kmpspMpc and a $\Lambda$ cosmology, 
$\Omega_M=0.3$, $\Omega_{\Lambda}=0.7$, $H_o=65$ \kmpspMpc
respectively)
at least one massive burst of star-formation has
already taken place, producing dust, then it is clear that it is more 
efficient with the current generation of radio telescopes to search 
first for the continuum dust emission from an object, following up 
secure detections with  searches for CO. 

\begin{figure}
\centering
\psfig{file=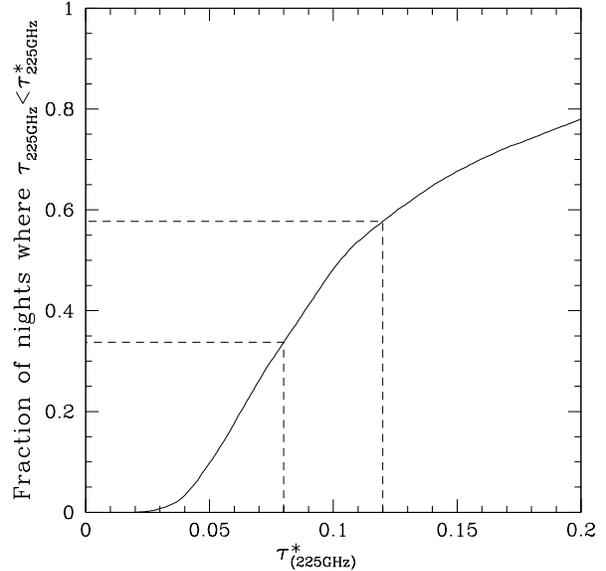, width=80mm}
\caption{Plotted are the fraction of nights for the first 9 months of 2000
for which $\tau_{\rm 225GHz} 
\leq \tau^{\ast}_{\rm 225GHz}$. Note that the abscissa extends to 0.2 only
(and hence the maximum fraction of less than 1) as this represents
the maximum $\tau_{\rm 225GHz}$ for which $850\mu m$ observations of 
the brightest sources can sensibly be made at 
JCMT (a zenith transmission  $>45\%$ ).
Tau measurements were made using the 
225 GHz CSO tipping radiometer. Data have been taken from the CSO 
web-page, kindly provided by Ruisheng Peng. 
The dashed lines delineate the range of tau values during which the 
sky transmission was between $\sim60\% - 70\% $, a $\tau_{\rm 225 GHz}$
range of 0.08--0.12 (referred to as ``grade 3'' conditions at the JCMT).  
Such conditions occur for approximately 25\% of the available night-time
observing time. Note, the term ``grade 1'' is used to refer to occasions
when the $850\mu m$ zenith sky transmission is better than about
80\% ($\tau_{225} < 0.05$), whilst ``grade 2'' conditions are those for which 
the zenith transmission is between about 70\% and 80\%. }\label{fig_tau_cumm}
\end{figure}

\begin{figure*}
\centering
\psfig{figure=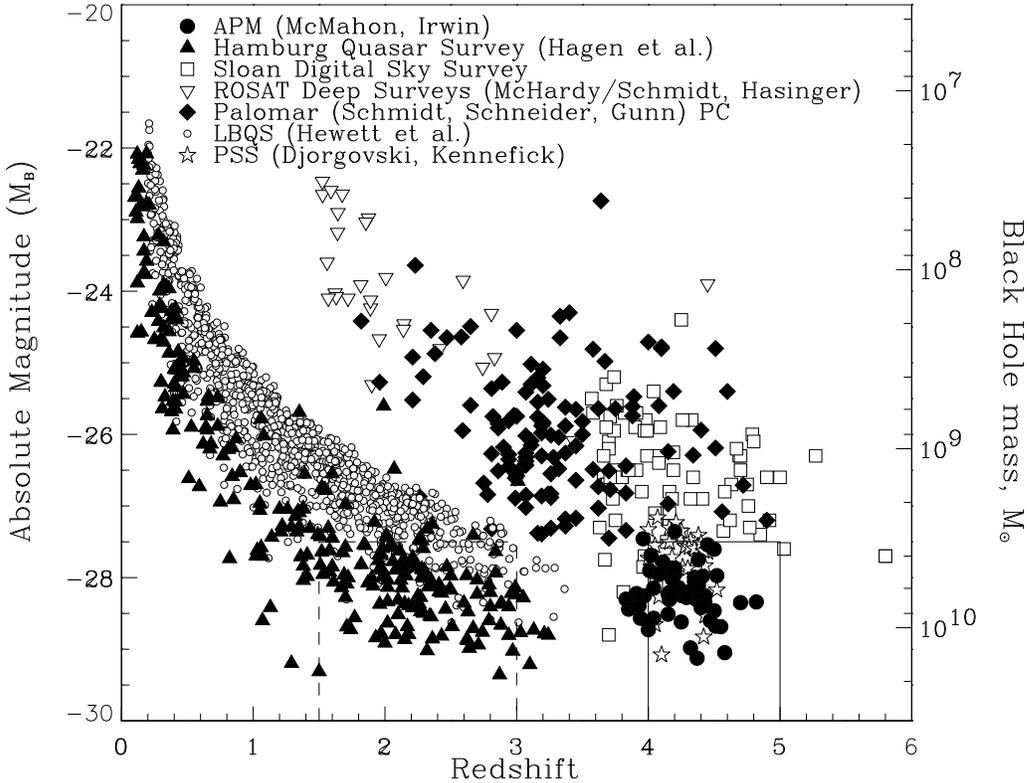,width=150mm}
\caption[]{Hubble diagram: absolute magnitude vs. redshift, of 
all known z$\ga$4 radio-quiet quasars and a selection of $z<3$ surveys.
Absolute magnitudes have been calculated using
a cosmology of $H_o=50$ \kmpspMpc, $q_o = 0.5$, and magnitudes 
taken from the literature. 
The sources delineated by the solid box are those that define the
optically-bright, z$\ga$4 parent sample of the survey described in this paper, 
whilst the sources delineated by the dashed box are those of the 
follow-up $z\sim2$ sample. Black-hole masses have been calculated using 
the relationship given in 
Eq. \ref{eqn_eddlum}, and a bolometric correction $L_{bol}/L_B = 12$
(Elvis et al. 1994)}\label{fig_hubble_diagram}
\end{figure*}

A $3\sigma$ detection of a source 
with $S_{850\mu m}=10$mJy takes approximately 25 minutes 
(excluding pointing and 
calibration overheads) using SCUBA/JCMT
under observing conditions for which the sky transparency is c. 65\%.  
Such conditions occur $\sim25\%$ of the time (Figure \ref{fig_tau_cumm})
and thus a survey for submillimetre-loud object at high redshift can slot 
well into periods during which the sky transmission is not high 
enough to undertake the high-priority projects requiring the
highest possible sky transmission. Indeed, under the very best 
observing conditions, when the zenith transparency at $850\mu m$ can be as 
high as 90\% , the required sensitivity can be reached in an 
elapsed time of around 12 minutes. Such observations become an inefficient
use of telescope time since they become dominated by slew
time and pointing checks. In contrast, the detection of associated molecular 
gas for a $10$mJy source would currently
take some tens of hours using the PdB Interferometer 
and the Owens Valley Millimeter Interferometer
\footnote{If we assume that the submillimetre flux is thermal in origin,
then with assumptions about the physical properties of the dust: 
$T_d = 40K$, a dust spectral index of $\beta = 2$ (see 
Priddey \& McMahon \shortcite{Priddey01} for 
a ``fiducial'' $z\ga4$ quasar), a power-law dust emission given by Eq. 
\ref{eqn_kappa}, we derive a dust-mass estimate of around $10^9$\Msun 
for $S_{850\mu m}= 10$mJy. Assuming a ``typical'' gas-to-dust ratio 
of 500, and $M_{H_2}/ CO$ conversion factor of $\alpha = 4$, the 
equivalent $CO$ gas mass that one might expect is $\sim$few x$10^{11}$\Msun
detectable at millimetre wavelengths using either the Plateau de Bure 
interferometer or the Owens Valley Radio Observatory Interferometer 
in some 10s of hours}.

Samples of statistically significant size over a wide range of redshifts
are crucial not only to studies of starformation and AGN activity 
at different epochs, but also to the determination of the role of 
starformation activity and black-hole accretion in the evolution of 
the Universe.

In this paper we report on the first results of an $850\mu m$ 
program to search for massive star-bursts in the host galaxies of the 
most optically bright radio-quiet quasars at $z\ga4$
using SCUBA at the JCMT. The aim of the survey is to 
identify a sample of $\sim$10 submillimetre-bright sources, from which  
it should in principle be possible to detect molecular gas 
emission using the current 
generation of millimetre interferometers. Based on extrapolations to 
$850\mu m$ of observations of radio-quiet quasars at $z\ga4$ 
by Omont {\it et al.} \shortcite{Omont96b}, 
we might expect to detect $\sim0.3$ 
submillimetre-bright sources per hour ($S_{850\mu m}\ga10$mJy) under 
modest submillimetre-observing conditions (sky transparency of 
c. 65\% at $850\mu m$). We note that this detection rate is based on the
simplifying assumption that the sample of 
Omont {\it et al.}\shortcite{Omont96b} has the 
same distribution of source properties as any sample that we might define 
for the purposes of these observations. Such newly identified 
submillimetre-bright sources, 
together with the 4 $z\sim4$ well-studied objects from which 
CO has already been detected, and those for which upper limits to a 
gas mass have been obtained, will provide a sample of statistically 
significant size ($N>3\sqrt{N}$), 
with which it will be possible to begin to draw conclusions 
about massive star-formation and its role in the high redshift universe.

We present here the interim results of our survey for a subset of 
38 sources and discuss the 
implications of these current and future observations 
on studies of star-formation at high-redshift. In a companion paper,
Omont {\it et al.} \shortcite{Omont01} present 1.2mm observations of 
the same parent sample of z$\ga$4 quasars. In
a future paper(Priddey {\it et al.}, in prep.) 
we will present an analysis of the combined sample of more than
100 z$\ga$4 quasars, 
including observations by others of different samples ,  
which have now been observed at either $850\mu m$ or 1.2mm, 

We have adopted an Einstein de Sitter cosmology of 
$H_o=50$ \kmpspMpc, $\Omega=0$ and, where useful,
include a comparison with parameters derived using a $\Lambda$-cosmology with 
$H_o=65$ \kmpspMpc, $\Omega_M = 0.3$; $\Omega_{\Lambda}=0.7$. 

\section{Sample definition}
\subsection{Optical properties}
The parent observational sample has been selected to include the most luminous
(rest-frame UV) $z\ga4$ radio-quiet quasars known as of January, 2000.
The quasar sample was primarily drawn from the APM multicolour
photographic surveys based on APM scans of UKST plates by
Irwin, McMahon \& Hazard(1991), Storrie-Lombardi {\it et al.}(1996, 2001) and
the Caltech DPOSS based on STSCI scans of POSS-II plates by 
Kennifick {\it et al.}(1995). The first step in the sample definition was to 
calculate the absolute magnitude in a consistent manner for all
($\sim$200) known z$\ga$4 quasars in order to select a sample 
on the basis of their rest-frame absolute B band magnitudes
(see Figure \ref{fig_hubble_diagram}).

Absolute $B$ band magnitudes ($M_{B}$) were evaluated using observed 
$R$ band magnitudes for all the quasars, assuming a spectral index
$\alpha=-0.5$ and using the relationship 
\begin{equation}
M_B =  R - dm(z) + 2.5log(1+z) +k_R(z) -0.5,
\end{equation}
where $R$ is the observed $R$-band($\lambda \sim6500\AA$)
magnitude and $dm(z)$ is the distance modulus, evaluated at the source 
redshift. 
$R$ band magnitudes were determined from APM scans of UKST $OR$ 
(5900--6900\AA) or of POSS-1 $E$ (6200--6800\AA) plates
from APMCAT (www.ast.cam.ac.uk/$\sim$apmcat;
McMahon \& Irwin(1992)). Errors on these magnitudes are estimated
to be $\sigma_R\sim0.3$. 
At $z\sim4$, the $R$ band starts to sample the rest-frame 
spectrum near the prominent Lyman-$\alpha$ emission line; and 
at higher redshifts, 
the band suffers from the strong intergalactic absorption shortwards
of Lyman-$\alpha$. To compensate for both of these effects, 
a redshift- and filter-dependent correction 
factor ($K_{R}(z)$) is invoked. 
Initially, for the purposes of sample selection,
we adopted the empirical correction of Kennefick, 
Djorgovski \& Meylan (1996). 
However, for the subsequent analysis, we derived our own correction,
based on the mean LBQS quasar spectrum \cite{Francis91}, combined
with the model of the hydrogen opacity of the intergalactic medium 
of Madau \shortcite{Madau95}, to obtain an average 
correction for each filter as
a function of redshift.
The correction was compared with empirical values derived from
all $z>4$ quasars for which we
possess accurately flux-calibrated spectra 
(see Storrie-Lombardi et al., 1996, and Peroux et al., 2001),
and is found to follow the trend of the data well. 
The scatter superimposed on this trend is large,
however, leading to an overall uncertainty in 
the magnitude $\sigma_M\sim0.5$. 
This scatter reflects not only the idiosyncratic
variations in the effects of emission and absorption features, but also
an intrinsic dispersion in the slope of the quasar continuum.
For some of the quasars in the sample, a more direct measure of
the continuum luminosity is available either through optical
spectroscopy or through photometry in the $K$ band,
which, at $z=4$, samples the rest-frame $B$ band directly.
However, for consistency and for the purposes of comparison, 
we use the uniformly available $R$ magnitudes to derive a homogeneous
measure of luminosity. Issues pertaining to the derivation of accurate 
source magnitudes will be addressed more completely in a following paper.

\begin{table*}
\begin{minipage}{150mm}
\caption[]{Optically and X-ray selected $z\ga4$ quasars with 1.4GHz 
associations in NVSS.
(Assuming a cosmology $\rm H_o=50$ \kmpspMpc, $q_0=0.5$,
and a spectral index for both radio and optical $\alpha=-0.5$)}

\begin{center}
\begin{tabular}{lccccrrcccc}
Quasar name & RA (J2000) & Dec (J2000) & $z$ & $dr$\footnote{Separation 
between optical and nearest radio source positions}   
& $\log{P(r)}$ \footnote{Logarithm of the probability that 
the optically selected quasar has been {\it falsely} associated 
with a radio source from the NVSS catalogue, separated by  distance $r$} 
& $S_{1.4}$ & $S_{850\mu m}$ \footnote{Extrapolated contribution by synchrotron
emission to any observed 850$\mu m$ flux, based on a spectral 
index, $\alpha$, of 0.7, $S_{\nu}\propto \nu^{-\alpha}$} & $M_B$ 
& $\log{L_{1.4}}$ \footnote{[$L_{1.4}$] = $\rm erg s^{-1} Hz^{-1}$}
 & $\log{R_{1.4}}$ \footnote{Ratio of radio (1.4 GHz) to optical (rest-frame 
B-band) luminosities} \\
&&&&\arcsec&(false)&mJy&mJy&&&\\
\hline
PSS J0121+0347$^{A}$   & 01 21 26.1 & +03 47 07   & 4.13 & 1.7 & $-$4.5& 78.6 & 1.65  & $-$27.5 & 34.69 & 2.97\\
PSS J0211+1107    & 02 11 20.0 & +11 07 17   & 3.99 & 0.5 & $-$5.5 & 44.1 & 0.93  & $-$27.7 & 34.41 & 2.61\\
PSS J0439$-$0207  & 04 39 23.2 & $-$02 07 02 & 4.40 & 0.5 & $-$5.5 & 43.8 & 0.92  & $-$27.5 & 34.49 & 2.77\\
PSS J2256+3230    & 22 56 10.4 & +32 30 19   & 4.04 & 2.41 & $-$4.1 & 14.7&0.31 & $-$26.0 & 33.95 & 2.83\\
\hline
BRI B0151$-$0025$^{A}$  & 01 53 39.6 & $-$00 11 05 & 4.20 & 7.3 & $-$3.2 & 3.0&  0.07 & $-$27.5 & 33.28 & 1.56\\
BR J0234$-$1806& 02 34 55.1 & $-$18 06 09 & 4.30 & 0.7 & $-$5.2 & 35.8&  0.75  & $-$27.5 & 34.38 & 2.66\\
BR J0324$-$2918& 03 24 44.3 & $-$29 18 21 & 4.62 & 0.6 & $-$5.3 & 236.8 &  5.00  & $-$28.8 & 35.26 & 3.02\\
BR J0355$-$3811& 03 55 04.9 & $-$38 11 42 & 4.58 & 13.4 & $-$2.6 & 2.4   &  0.05  &$-$29.4 & 33.26 & 0.78\\
BR J0523$-$3345& 05 25 06.2 & $-$33 43 06 & 4.40 & 1.0 & $-$4.9 & 188.7 & 3.96   &$-$28.1 & 35.12 & 3.16\\
BR J1053$-$0016 & 10 53 20.4 & $-$00 16 49 & 4.29 & 2.6 & $-$4.1 & 9.7& 0.2 &$-$27.0 & 33.81 & 2.29\\
BR J1305$-$1420   & 13 05 25.2 & $-$14 20 42 & 4.04 & 0.96 & $-$4.9&  20.6&0.43  & $-$28.0 & 34.09 & 2.17\\
\hline
SDSSp J0131+0052$^{B}$   & 01 31 08.2 & +00 52 48   & 4.19 & 3.76 & $-$3.8 & 4.2 &0.09 & $-$25.8 & 33.43 & 2.39 \\
SDSSp J0210$-$0018$^{B}$ & 02 10 43.2 & $-$00 18 18 & 4.77 & 3.24 & $-$3.9 & 11.5 & 0.24 & $-$26.8& 33.97 & 2.53 \\
SDSSp J0300+0032$^{B}$   & 03 00 25.2 & +00 32 24   & 4.19 & 7.80 & $-$3.1 & 7.1 & 0.15& $-$26.0 & 33.70 & 2.58 \\
SDSSp J1235$-$0003$^{C}$ & 12 35 03.0  & $-$00 03 31.8 & 4.69 & 0.48 & $-$5.5 & 19.5& 0.41 & $-$26.3& 34.19 & 2.95 \\
\hline
PC J0027+0525$^{A}$     & 00 29 49.97 & +05 42 04.4   & 4.10 & 8.47 & $-$3.0 & 4.8 & 0.10 & $-$24.8 & 33.47 & 2.83 \\
PC J2331+0216$^{A}$     & 23 34 32.0 & +02 33 22   & 4.09 & 4.92 & $-$3.5 & 2.7 & 0.06  & $-$27.3 & 33.22 & 1.58\\
\hline
\end{tabular}
\end{center}
\label{tab:zgt4_rloud}
For completeness all quasars in our parent sample for which 
potential radio counterparts were found are listed in the table. 
Those marked with $^{A}$ and $^{B}$ have already been found by 
Stern {\it et al.}, \shortcite{VLAsource} and Fan {\it et al.}, 
\shortcite{Fan01} to have NVSS counterparts, whilst the source
labelled $^C$ has a radio-counterpart \cite{Fan00} as found in the FIRST 
survey \cite{Becker95} 
\\ \end{minipage}
\end{table*}

Objects already observed to a sensitivity equivalent or better than that
of the survey as parts of other 
(sub)millimetre programs were excluded from the target list
(eg. the 5 quasars observed by McMahon etal, 1999;
BR B0019$-$1522, PSS J0134$+$3307, PSS J0747$+$4434,
BR B1600$+$0729, BR B2237$-$0607), as were sources with transit 
elevations of less than 60 degrees.
An exception to this was PSS J1048$+$4407,  previously observed
and detected at $1.25$mm at the IRAM 30-m (Maoli {\it et al.}, in prep),
but not confirmed with 1.35mm observations at the 
Plateau de Bure interferometer \cite{Guilloteau99}.

\subsection{Radio properties} 
The final source selection criterion was that of radio flux.
We have used the NRAO VLA Survey Survey(NVSS) to measure or 
place upper limits on the radio emission from all z$\ga$4 quasars that
were potentially in our sample.
The NVSS survey \cite{Condon98} is a sensitive radio survey that was 
carried out with the VLA at 20cm in B-array,  
with a 5$\sigma$ sensitivity of $\sim2.5$mJy/beam. To determine 
which of the optically selected quasars are radio-loud, we 
cross-correlated the optical positions of our parent sample with 
sources in the NVSS catalogue. To assess the probability 
of a radio source being matched to one of the optically selected quasars
by chance, the background source density was determined by
counting the density of radio sources in the annulus with radius
100\arcsec to 1000\arcsec, resulting in a value of 50 deg$^{-2}$. 
The rms positional error of the NVSS catalogue is given as 7\arcsec
and thus, if one assumes
that any radio source within 21\arcsec is a potential match, one expects 
0.005 chance associations per source, or a probability of 1 in 200 of 
incorrectly attributing radio emission to an optically selected 
quasar. Even if this radio 
emission were not to be from the quasar itself, a 
fraction of the emission would still be picked up by JCMT/IRAM main and 
error beams.

Listed in Table~\ref{tab:zgt4_rloud} are all optically-selected 
z$\ga$ 4 quasars for which
we find an NVSS source within a 30\arcsec radius -- a total of 17 objects --
along with their derived parameters.  A detailed discussion of the radio
properties of the sample is beyond the scope of
this paper and will be presented elsewhere. 
All sources with $S_{1.4GHz} > 2.5$mJy were excluded from the final SBQS
target list. Based on a 
flux extrapolation of $S\propto\nu^{-\alpha}$, and a    
canonical spectral index of $\alpha=0.7$, we therefore 
expect the maximum synchrotron contamination at $850\mu m$ to be less than 
$0.1$mJy. Thus, it is reasonable to assume that the emission from 
any source that is detected at $850\mu m$ is likely to be 
thermal in origin. 

\medskip
Our final target list comprised 76 sources with 
$M_B < -27.5$, augmented by 22 sources 
with $-27.0 < M_B < -27.5$ which were included to improve sky coverage: 
15 quasars were taken from the APM B/R and B/R/I survey (BR and BRI sources 
: \cite{Irwin91,SMIH96,SIMH01}), 70 from the Palomar Sky Survey 
(PSS sources: \cite{Djorgovski00}), 
2 from the Palomar Survey (PC sources : \cite{Schneider91a,Schneider91b}),
1 from the VLA (VLA source: \cite{VLAsource}) and 
9 from the Sloan Digital Sky Survey (SDSSp sources: 
Fan {\it et al.}\shortcite{Fan99,Fan00}).

\section{Observations and Data Reduction}
\subsection{Observations}
Observations were made using the wide-band $850\mu m$ filter on SCUBA, 
during periods when the zenith sky transparency was 
$\sim60-75\%$  ($\tau_{850\mu m}\sim0.08 - \tau_{850\mu m}\sim0.12$, 
referred to as grade 3 weather in JCMT parlance). Sources were observed by the 
``displaced observer'' with priority set by source availability and 
luminosity. The standard $850\mu m$ photometry observing mode was used --  
a 9-position jiggle superposed onto a 7 Hz/60\arcsec chop throw of 
the secondary mirror in the azimuthal direction,   
implemented to remove sky emission offsets.
Gradients in sky emission were taken out by ``nodding'' the telescope 
-- a physical movement of the telescope to the off-position every 18s. 
A typical observation consisted of sets of c. 30--50 18-second integrations,
with the total number of integrations chosen to achieve a sensitivity of  
$S_{850\mu m}(1\sigma)\sim 3.3$mJy. Pointing was checked regularly, and 
found to be accurate to better than a few arc-seconds. Astronomical 
seeing was monitored using the 5GHz CfA phase interferometer 
-- the phase monitor was not in continual use however, when working, the 
data suggested that the seeing was not worse than 3\arcsec
\footnote{ We note that in the worst case, the combined pointing and 
seeing errors of  3\arcsec would result in a loss of point-source coupling of 
40\%, assuming a Gaussian beam profile with a FWHM of 14\arcsec}.
Sky transparency was monitored using 
regular $850\mu m$ sky-dips and data taken by the 225 GHz Caltech Submillimetre-Observatory  
(CSO) tau-meter. Flux calibration was achieved using the 
primary calibrators 
Mars and Uranus, as well as the secondary calibrators HLTau, IRC10216 
\footnote{IRC10216 was adopted as calibrator on one
occasion only, due to its periodic variability. When used as a calibrator, 
the output flux of IRC10216 was at a maximum}, OH231 and 16293
\footnote{ See the JCMT web-page for the most
recent calibrator fluxes}. Absolute fluxes for the planets were 
derived using the Starlink package, FLUXES. The 
average time to complete the observation of a single source, including 
pointing, calibration, sky-dip and slewing overheads, was over 1 hour. 
This is quite a bit longer than would be the case for a contiguous set 
of observations, reflecting the start/stop nature of a back-up program. 

Here, we discuss observations of a subset of the 
target list: sources observed during the period February -- July, 2000.    

\subsection{Data reduction}
Data were reduced using the Starlink SURF reduction package \cite{Lightfoot}, 
using the automated ORACDR pipeline \cite{Jenness} as well as a 
basic SURF-based pipeline written in-house prior to the release of ORACDR.     
Two different schemes were implemented to determine an appropriate 
measure of the background sky-noise, the first using the mean of the 
inner ring of six pixels surrounding the central on-source bolometer 
to define the temporal sky variations, and the second the median of 
all bolometers. In general, only small differences in the 
final fluxes and rms noise determinations were seen between the two 
methods for observations in which detections were made. There were, 
however, more significant differences in rms values determined 
for non-detections. In particular, the sky-noise correction 
derived from the inner-ring mean resulted in a non-zero offset in the 
distribution of the signal/noise ratios for all off-source bolometers, 
indicating incomplete sky-removal. This was in contrast to the median-derived
sky correction which we chose to adopt in the final analysis.
All quasar data were clipped at the $20\sigma$ level prior
to applying the sky-subtraction in order to eliminate transient spikes.  
Corrections for atmospheric extinction were made using sky opacities 
derived from 850$\mu m$ sky-dips. 
Individual sets of integrations were gain-calibrated and then 
concatenated, with a $3\sigma$ despiking applied to remove 
outlier data points. \newline
\begin{figure}
\centering
\psfig{file=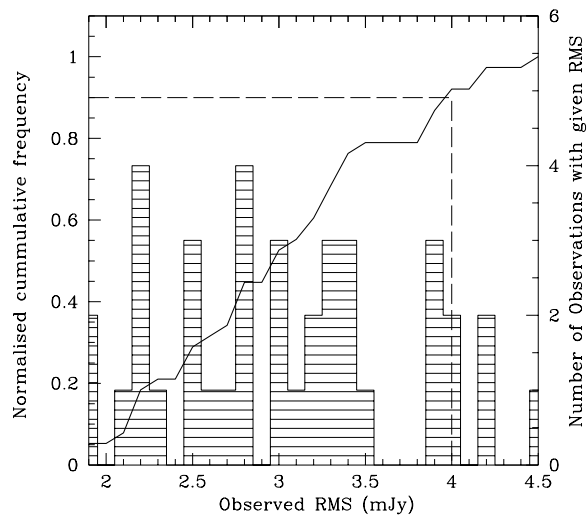,width=80mm}
\caption{
A plot of the rms values obtained for all observations reported in this   
paper. The histogram denotes the number of sources with a given rms, while
the solid curve denotes the normalised cumulative number of sources with
a given or lower rms. Note the dashed line delineates 90\% of the objects     
observed, and defines the cutoff rms chosen for the statistical sample
discussed in the text.
}\label{fig_sample_defn}
\end{figure}

\begin{table*}
\begin{minipage}{150mm}
 \caption[]{Observational parameters of the observed z$\ga$4 sample}
  \begin{tabular}{@{}lllllllllll@{}}
\footnote{BR/BRI source positions taken from the APM catalogue; PSS source
parameters have been taken from the Palomar Sky Survey
(\cite{Djorgovski00})
SDSSp source positions have been taken from the
Sloan Digital Sky Survey (\cite{Fan99,Fan00})} &
$z$ \footnote{Palomar Sky Survey redshifts have been taken from 
Djorgovski \shortcite{Djorgovski00}, with the exception of sources marked $\ast$, where
redshifts have been taken from Peroux {\it et al.} and
Storrie-Lombardi {\it et al.}} & RA
& Dec.  & $R$ \footnote {$R$ magnitudes determined from APM
scans of the UKST $OR$ (5900--6900\AA) or POSS-1 $E$ (6200--6800\AA) 
plates. Errors are estimated to be $\sigma_R\sim0.3$} 
 & $M_B$ \footnote{See text for an explanation of
the method used to evaluate the absolute source magnitudes}  & obs. & integ. 
\footnote{Integration times have been calculated based on 
no: integrations x 18s/integration x 1.5, where both chopping and a 
50\% observational SCUBA overhead 
has been allowed for,  This {\it does not include} any overhead for calibration or pointing.} &
 Flux \footnote{Where a $3\sigma$ detection was not been achieved, the
tabulated flux is given as the on-source signal only} & rms & S/N\\
  & & (J2000)  & (J2000)  & & & date & time (s) & (mJy) & (mJy) & \\

\hline
PSS J0452+0355 & 4.38& 04 52 51.5& +03 55 57& 19.1&$-$27.6 & 7Feb00& 2640 & 
10.6&2.1& 5.0 \\
PSS J0808+5215 &4.45  & 08 08 49.4 & +52 15 15 & 18.3&$-$28.7&4Feb00& 2640  & 17.4&
2.8&6.2 \\
PSS J1048+4407 
\footnote{ The position reported here is that taken from
the Djorgovski web page. A check of the position using the APM catalogue
{\it after} the JCMT SCUBA observations had been made revealed that the published position is out by 2\arcsec.
The correct position is: (J2000) 10 48 46.64 +44 07 10.9}
&4.38*& 10 48 46.6 & +44 07 13 &19.6
&$-$27.4&12Feb00/3Jul00& 3940  & 12.0 & 2.2 & 5.3\\
PSS J1057+4555 &4.12& 10 57 56.3 & +45 55 53 & 16.5&$-$29.4 & 11Feb00 &2640  & 19.2 &
2.8&6.9 \\
PSS J1248+3110 &4.35 &12 48 20.2 & +31 10 44 & 18.9& $-$27.6 & 22Apr00& 1340& 12.7
&3.4&3.8 \\
PSS J1418+4449 &4.28 & 14 18 31.7&+44 49 38& 17.6&$-$28.6&23Apr/3Jul00& 4200  
&10.4&2.3&4.5 \\
PSS J1646+5514
\footnote{The flux reported is the weighted sum of 4 sets of integrations
that differed considerably in statistical significance, ranging from
$3\sigma$ and better detections to a near-zero on-source flux.}
 &4.04* &16 46 56.5    & +55 14 46 &  17.1&$-$28.7 &13/14.02/7Feb00& 5240 & 9.5 &2.5
&3.8\\
PSS J2322+1944& 4.17 & 23 22 07.2&+19 44 23& 17.7 &$-$28.1 &27Jul00& 1600 & 22.5 &2.5
&8.9 \\
               &            &              &         &
&             &&& \\
\hline
               &            &              &         &
&             &&& \\
PSS J0007+2417 &4.05&00 07 38.7&+24 17 24& 18.3&$-$27.6 & 13Jun00 & 1340 & 1.2 &3.5 &
0.3\\
PSS J0014+3032 &4.47& 00 14 43.0 & +30 32 03 & 18.7&$-$28.1&30May00 & 1340 & 10.6 &
4.2 & 2.5\\
PSS J0133+0400 &4.15*& 01 33 40.3&+04 00 59 & 18.3&$-$27.6 &13Jun00 & 1340  & $-$1.0
&2.8 &$-$0.4\\
BR\phantom{A}B0300$-$0207 &4.25 &03 02 53.0 &$-$01 56 07& 18.5&$-$27.6
&12Feb00 & 3940 & $-$1.7& 2.8&$-$0.6\\
BR\phantom{A}B0351$-$1034 &4.35* & 03 53 46.9 & $-$10 25 19& 18.6&$-$27.8&
12Feb00& 2640  & $-$2.8&3.0&$-$0.9\\
BR\phantom{A}B0401$-$1711 &4.24& 04 03 56.6 & $-$17 03 24  & 18.7&$-$27.6
&11Feb00 & 3940  &2.6& 3.4&0.8\\
PSS J0852+5045 & 4.2&08 52 27.3&  +50 45 11  & 17.7  &$-$28.2&9/12Feb00&5240 & 
1.8&2.2&0.8\\
PSS J0926+3055 &4.19 &09 26 36.3& +30 55 05& 16.4 &$-$29.5& 4/11Feb00& 6540& 0.8&
1.9& 0.4\\
BRI B0945$-$0411&4.14 &09 47 49.6  & $-$04 25 15 & 18.8 &$-$27.3&5Feb00&2640  & 2.1&
2.2&1.0\\
PSS J0957+3308 &4.25& 09 57 44.5 & +33 08 20 & 17.8&$-$28.3
&11Feb/7Mar00& 5396 & 3.7&1.9& 2.0 \\
BRI B1013+0035 &4.40 &10 15 49.0 & +00 20 19  &  18.8&$-$27.9&12Feb00 & 2640& 5.5&
2.7& 2.0\\
PSS J1026+3828 &4.18 &10 26 56.7  &  +38 28 45 & 18.1&$-$27.8&12Feb00 & 2640& 5.2&
2.5&2.1\\
PSS J1058+1245 &4.33 &10 58 58.4 & +12 45 55 & 17.6&$-$28.8&12Feb00 &2640& 0.00 &
2.2&0.0 \\
BRI B1110+0106 &3.92* &11 12 46.3 & +00 49 58& 18.3&$-$27.1&5Apr00& 1340 & 7.1& 3.4&
2.1\\
SDSSp J1226+0059&4.25 &12 26 00.7   & +00 59 24 & 18.9&$-$27.0 &14Feb00 & 2640 & 
$-$0.06 & 3.3&0.0\\
BR\phantom{A}B1302$-$1404 &4.00*& 13 05 25.2&$-$14 20 42& 18.6&$-$27.4&
23Apr00 & 1600 & 1.0& 3.2&0.3\\
PSS J1315+2924  & 4.18 &  13 15 39.8 & +29 24 39& 18.5&
$-$27.4&22Apr00& 1340 & 4.1&3.3&1.3 \\
PSS J1326+0743 & 4.17&13 26 11.8 & +07 43 58& 17.5&$-$28.4&3Jul00& 1340 & $-$0.3 &
3.9 &$-$0.1 \\
PSS J1347+4956 &4.46&13 47 43.3& +49 56 21& 18.5&$-$28.3&22Apr00 & 1470 & 8.5 &
3.9&2.2 \\
SDSSp J1413$-$0049&4.14 &14 13 32.4&$-$00 49 10& 19.1&$-$27.0 &14Feb00&2640&
$-$1.0&3.2&$-$0.3\\
PSS J1432+3940&4.28 &14 32 24.8 &+39 40 24& 18.0&$-$28.2&3Jul00 & 2640&  $-$1.4&3.3
& $-$0.4\\
BR\phantom{A}B1500+0824  &3.94*& 15 02 45.4 & +08 13 06 & 19.3&
$-$26.3&14Feb00 & 3940 & 6.2& 3.0&2.1\\
PSS J2154+0335 &4.36 &21 54 06.7 &+03 35 39& 19.0&$-$27.5&5Apr00&1860& 2.6&
2.6&1.0 \\
PSS J2155+1358 &4.26*&21 55 02.2&+13 58 26 & 18.0&$-$28.1&5Apr00 & 1860 & 0.8&
3.1&0.4\\
PSS J2203+1824 &4.38 & 22 03 43.4& +18 28 13& 18.0&$-$28.7&30May00 & 1340 & 3.0&
4.2&0.7\\
BR\phantom{A}B2212$-$1626&3.99*&22 15 27.3& $-$16 11 33&18.1 &$-$27.9&
30May00& 1340 & 7.9 & 4.5&1.7\\
PSS J2238+2603 &4.03& 22 38 41.6& +26 03 45 & 17.0&$-$28.9 &30May00 & 1340 & 2.2&
3.9&0.6\\
PSS J2241+1352 &4.44*& 22 41 47.8& +13 52 02& 19.1&$-$28.0&5Apr/30May00&2250 &
$-$0.2& 3.0&$-$0.05\\
PSS J2323+2758 & 4.18 & 23 23 40.9 & +27 58 00 & 18.6&$-$27.3&30May00& 1340 &  $-$2.8
&4.0&$-$0.7\\
PSS J2344+0342 & 4.24*& 23 44 03.2& +03 42 26& 18.2&$-$28.1&30May00& 1340 & 11.3 &
4.0&2.8\\[10pt]
\hline
\end{tabular}
\end{minipage}
\end{table*}

\section{Results}
Source positions, redshifts and derived absolute 
B-magnitudes ($M_B$) are tabulated in Table 2, along with  
dates of observation, observed flux densities, 
rms values and signal-to-noise ratios.
Note that calibration errors of $\sim$20\% have not been included in the 
error budget.

A total of \obs quasars have been observed (see Table 2):  
\det sources were detected with $S_{850\mu m}\ga10$ mJy, at 
statistical significances of better than $3\sigma$. 30 sources were 
observed but not detected:   \less_10 sources   
with $3\sigma < 10$ mJy, 8 sources  
with $10mJy < 3\sigma < 12$ mJy and \great_12 sources 
with $3\sigma > 12$ mJy (see Table 2). Achieving the pre-requisite
survey sensitivity consistently proved to be difficult -- the initial
$3\sigma\sim10$ mJy sensitivity threshold was frequently both bettered 
(\less_10 non-detections) and not reached (\total_great_10 non-detections).
\newline

The size of the observed sample is large enough not only for 
a significant number of submillimetre-bright sources to have been 
newly identified, but also
to define a sample with which to start to assess the detection rate as well 
as detectability as a function of redshift and absolute magnitude. Here
we adopt a $3\sigma\sim$12mJy cut-off (a limit which includes 90\% of sources
observed in our program, Figure \ref{fig_sample_defn}), 
noting that there are possibly as many as 3 
sources which may have $S_{850\mu m}\sim10$mJy, but for which the 
observations obtained were not deep enough to obtain a $3\sigma$ detection. 
Sources were initially included in our target list based on 
an {\it approximate} $3\sigma\sim10$ mJy cutoff at $850\mu m$  -- 
BR B1500+0824 and BR B1302-1404 were thus included in the 
JCMT sample in spite of having been observed at $1.25$mm at 
IRAM-30m down to an equivalent $S_{850\mu m}\sim10$ mJy. 
We include these two sources in our statistical sample which includes
all sources for which the observed sensitivity is 
$3\sigma\leq 12$ mJy -- a total of \stat objects.

\section{Discussion}
\subsection{Newly detected submillimetre-bright sources}
To date, our survey has identified a total of \det submillimetre-bright 
sources (where here we include PSS J1048$+$4407)
with fluxes between $\sim10-25$ mJy. 
This equates to a \jcmt_det_rate detection rate, which is comparable to 
the rate of \omont_det_rate seen by Omont {\it et al.} \shortcite{Omont96b}
,and based on 1.25 mm observations of a sample of  radio-quiet, optically 
selected APM quasars with z$\ga$4.
We stress here that the comparison between the detection 
rates of the two different samples is valid ONLY if the underlying 
properties of the constituent sources are the same. We defer a more
detailed analysis of this issue to another paper     
\footnote{To make the comparison between the two different
samples, observed at two different wavelengths, we use the flux
ratio of $850\mu m/1250\mu m$ derived by Priddey \& McMahon 
\shortcite{Priddey01}
for z$\ga$4 quasars. At $z\sim4.2$ this ratio is $\sim2.6$, and thus 
we can compare the sample here with sources observed to $1\sigma\leq 1.5$ mJy 
at $1.25$m by Omont {\it et al.}\shortcite{Omont96b}. 
In evaluating our detection rate we have 
excluded BR B1500$+$0824 and BR B1302$-$1404 because of the common non-detection
to the IRAM-30m sample}. 
In each case, the measured fluxes are
more than a factor of 100 in excess of what might be expected based on 
the NVSS radio limits. Thus, the single submillimetre
flux measurements, combined with the a-priori radio limit of our parent sample,
suggest strongly that we are detecting submillimetre emission from these 
sources that is thermal in origin.

How do the fluxes reported here compare with those measured for 
other z$\ga$4 radio-quiet quasars? 
The brightest radio-quiet quasar z$\ga$4 host galaxy observed both at 
submillimetre and millimetre wavelengths remains BR B1202-0725 
(Isaak {\it et al.} \shortcite{Isaak94}, McMahon 
{\it et al.}\shortcite{McMahon94}, Buffey {\it et al.}, in prep) -- 
with an $850\mu m$ flux of almost greater than twice that of the brightest 
sources detected in this survey,
PSS J0808$+$5215, PSS J1057$+$4555 and PSS J2322$+$1944. 
Arc-second resolution ( c. 1-2\arcsec) imaging with the 
Plateau de Bure interferometer (PdB) by Omont {\it et al.}\shortcite{Omont96b}
have shown BR B1202-0725 to be a composite source, made up 
of two comparable 
gas-rich and possibly interacting galaxies.
If we assume that the $850\mu m$ flux distribution is similar to that of 
the 1.25mm flux, split equally between the two objects, then BR B1202$-$0725
is no longer such an exceptional object, and has a flux comparable to 
that of the brightest sources present here. With the further assumption that
PSS J0808$+$5215, PSS J1057$+$4555 and PSS J2322$+$1944 are all single sources,
there is a {\it suggestion} of a maximum observed flux of
$S_{850\mu m}^{max}\sim20$ mJy. The origin of such a maximum is
intriguing. A possibility is that the 
starformation is self-regulated, with the large number of supernovae events in 
the massive star-burst disrupting the local ISM and blocking  star-formation. 
We are currently obtaining both deep radio observations with the VLA 
and millimetre observations with PdB, with which we will be
able to establish the degree of extension of the sources. 

We note that the sources we detect here are as bright, if not brighter 
than those detected in blank-sky and cluster-lensing surveys. 
Interestingly, the two  
brightest sources detected in the cluster lensing surveys \cite{Ivison98,Knudsen00}
have $850\mu m$ fluxes of just over $S(850_{\mu m})\sim20mJy$, and are 
both AGN at $z\sim 2$. For comparison, 
the measured fluxes of the brightest and median source in the 
Canada-UK-Deep-Sky-Survey are $\sim8$mJy and $\sim4.5$mJy respectively 
(eg. \cite{Eales99}). The 200-square arcminute survey ``8mJy survey'' 
($3\sigma<$8mJy (Dunlop 2000, Scott {\it et al}., in prep.) 
has yielded 24 sources, detected at statistically 
significant levels, with $10$ brighter than $10$mJy, and the 
brightest with a flux of 
$15$mJy. We note here that others have also started similar surveys to 
our own at 1.25mm using the IRAM-30m \cite{Carilli01a}, but leave a
comparison between surveys to a companion paper. 

\begin{figure*}
\centering
\psfig{file=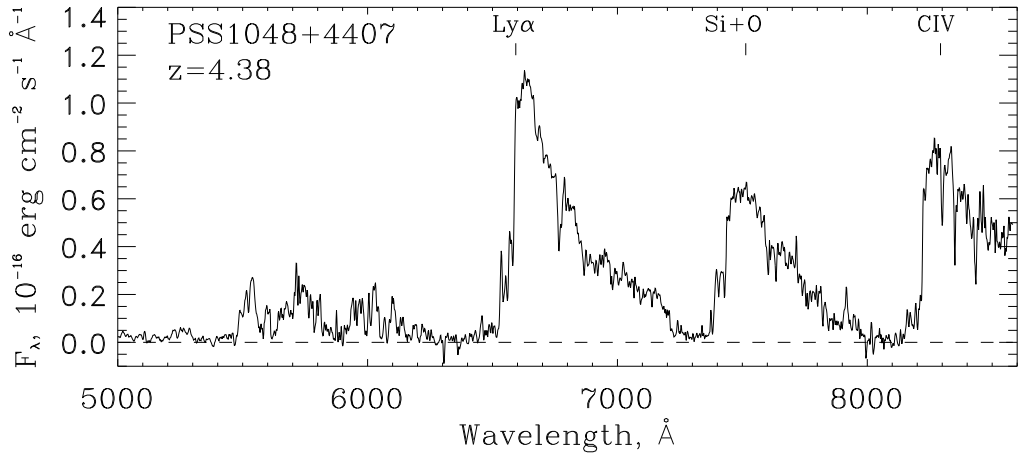, width=150mm} 
\caption{An optical spectrum of PSS J1048$+$4408, taken 
at a resolution 
of $5\AA$ using the ISIS spectrograph on the WHT by Peroux et al. . 
On the basis of this spectrum, we classify this quasar as a BAL, not 
possible with previous spectra because of the limited spectral coverage.
The presence of broad absorption lines makes an accurate determination of the 
host galaxy redshift difficult, however we derive redshifts of: $z=4.422$ 
($Ly\alpha(1216\AA)$, $z=4.367$ (Si+O $(1400\AA$) and $z=4.354$ 
(CIV($1539\AA$)).
From an average of these three lines we derive a source redshift of $z=4.38$, 
in contrast to $z=4.45$ as first derived by Kennefick et al. We note that even this 
redshift estimate could be biased high, as explained in the text}

\label{fig_spectrum_pss1048}    
\end{figure*}

\subsection{PSS J1048$+$4407}
In addition to identifying new submillimetre-bright sources, 
we detected PSS J1048$+$4407, a source for which observations at 
$\sim 1.3$ mm with IRAM-PdB and the IRAM-30m
were in conflict. 

\subsubsection{Single-dish vs. interferometric observations} As reported by \cite{Guilloteau99}, 
PSS J1048$+$4407 was not detected at 1.35mm using the PdB
interferometer ($S_{1.35mm} = 0.25 \pm 0.68$ mJy) in the CD configuration, 
but was tentatively detected by Maoli {\it et al} (in prep.)
1.25mm at the IRAM-30m ($S_{1.25 mm}\sim 3 \pm 1$mJy). Our 
result, $S_{850\mu m} = 12.0 \pm 3$mJy, is consistent, to
within the error-bars,  with the 30-m single-dish measurement if we assume a 
ratio of $S_{850\mu m}/ S_{1.25mm} \sim 2.5$, as derived by 
Priddey \& McMahon \shortcite{Priddey01}.
Neither of the two single-dish measurements, however, are consistent with the
interferometer observations. Pointing errors in the interferometer map 
and the effects of anomalous refraction would need to be much larger than 
realistically possible to explain the non-detection with the 
interferometer, given that at $\sim 220 $GHz the primary beam is $\sim 27$\arcsec. 
It is more likely that the host galaxy of PSS J1048$+$4407 is 
spatially extended on the scales greater than the synthesized 
beam of the interferometer ($\sim$ 3\arcsec x 2\arcsec).

To resolve out a substantial fraction of the millimetre flux, 
the spatial extent of the emission
does not need to be much greater than the size of the beam, depending 
more critically on the projected baselines/effective U-V coverage of the 
observations and the source extension relative to the position angle 
of the beam. At a redshift of $z\sim4.4$, 1\arcsec  equates to a linear size 
of c. 6kpc 
To resolve out 
emission over arc-second scales suggests that the star-burst activity,
as traced by the millimetre emission, is taking place over the 
whole galaxy rather than confined to nuclear regions as observed in 
Arp220 \cite{Downes98,Sakamoto99}, Arp193 and Mrk273 \cite{Downes98}.
This is in contrast to  
BRI B0952$-$0115 and BR B1335$-$0417 where a comparison between the 
single-dish and interferometric observations suggests that in each case
a significant fraction of the millimetre emission is unresolved, but 
similar to the case of BR1117$-$1329 and BR1144$-$0723, as discussed in 
Omont {\it et al.}, \shortcite{Omont01}.

\subsubsection{Updated redshift}
Also reported in Guilloteau {\it et al.}, \shortcite{Guilloteau99}
is the non-detection of CO (5-4) line 
emission using the PdB interferometer. This is interesting in light of the
$850\mu m$ detection presented here, as the source is one of two  
from five with a previously inferred $S_{850\mu m}>10$mJy for which 
a search for CO emission failed. 
A reassessment of the source redshift based on a new optical spectrum 
taken at the WHT by Peroux {\it et al.} \shortcite{Peroux01} 
(Figure \ref{fig_spectrum_pss1048})
reveals that the CO observations were centred at an erroneous redshift. 
Using the combination of the Ly$\alpha$, CIV and the SiIV+OIV emission lines, 
we derive a redshift of $z=4.38$, in contrast to that 
derived from the Ly$\alpha$ line only by Kennefick {\it et al.}\shortcite{Kennefick95},  $z=4.45$, and 
the redshift at which a search for CO emission was made 
\cite{Guilloteau99}.  The difference in redshift is significant -- 
much greater than that covered by the interferometer correlators. 
Our average redshift may, however, be biased high as it
contains the measurement based on the Lyman-$\alpha$ line which
is observed to be
systematically higher than the other two emission lines(CIV, SiIV+OIV).
It is possible that the Lyman-$\alpha$ line profile is affected
by absorption due to the NV(1240\AA) line which would bias
its peak redward. Using the CIV line alone, we 
derive a systemic redshift of $z=4.360$, correcting for the   
systemic shift from rest for the line as derived by
Tytler \& Fan (1992). 

Given the discrepancy between the observed
and newly derived redshift we suggest that the failure to detect CO emission 
is due to incorrect redshift, rather than an anomalous gas-to-dust ratio. 
This serves as a note of caution -- given the limited bandwidth 
of the current generation of millimetre and submillimetre telescopes, the 
non-detection  of any spectral line emission should be considered both in 
an astronomical context and, as important, in terms of a possible
error in redshift. 

\begin{table*}
\begin{minipage}{160mm}
  \caption[]{Derived properties of the detected (submillimetre-loud) z$\ga$4 
radio-quiet quasars sources -- see Section 6.3 for the definitions
used in deriving source properties}

  \begin{tabular}{@{}llllllllllll@{}}
\hline
Source name & $z$
& $M_B$& $\nu L_B$ & $M_{BH}$ & $\dot{M}_{\rm acc}$ & $M_{\rm sph}$ 
& $S_{850}$ & $M_{D}$ &  $L_{FIR}$ &  SFR & $\tau_*$\\
& & 
&$10^{13}\Lsun$&$10^9M_{\odot}$&$\frac{\rm\thinspace M_{\odot}}{\yr}$&
$10^{11}$\Msun 
& mJy& $10^9$\Msun & $10^{13}\Lsun$ & $\Psi\frac{\Msun}{\yr}$ 
& $\frac{\Gyr}{\Psi}$\\
(1)          &  (2)&  (3)  & (4) & (5)  & (6)& (7) &(8) &(9) & (10)&(11)&(12)\\
\hline
PSS J0452+0355&4.38&$-$27.6& 0.9 & 3.6  & 80 & 7.3 &10.6&0.9 & 1.2 & 1200&0.6\\
PSS J0808+5215&4.45&$-$28.7& 2.5 & 10.0 & 220& 20  &17.4&1.5 & 2.0 & 2000&1.0\\
PSS J1048+4407&4.38&$-$27.4& 0.8 & 3.0  & 70 & 6.0 &12.0&1.1 & 1.3 & 1300&0.5\\
PSS J1057+4555&4.12&$-$29.4& 2.8 & 19.0 & 420& 38  &19.2&1.8 & 2.3 & 2300&1.7\\
PSS J1248+3110&4.35&$-$27.6& 0.9 & 3.6  & 80 & 7.3 &12.7&1.1 & 1.4 & 1400&0.5\\
PSS J1418+4449&4.28&$-$28.6& 2.3 & 9.1  & 200& 18  &10.4&0.9 & 1.2 & 1200&1.5\\
PSS J1646+5514&4.35&$-$28.7& 2.5 & 10.0 & 220& 20  &9.45&0.8 & 1.1 & 1100&1.9\\
PSS J2322+1944&4.17&$-$28.1& 1.4 & 5.8  & 130& 12  &22.5&2.0 & 2.6 & 2600&0.4\\
\hline
median z$\ga$4 SBQS \footnote{The quantities and parameters evaluated for the 
median quasar are based on the median observables, from which parameters
are derived.} 
&4.35&$-$28.4 &1.8& 7.4& 165& 15&12.4&1.1&1.4& 1350&1.1\\ 
\footnote{Median quasar properties evaluated as above, based on a 
$\Lambda$ cosmology, $\Omega_{\Lambda}=0.7$, $\Omega_{M}=0.3$, 
and $H_o=65$\kmpspMpc. For comparison 
the ratio of the square of the luminositiy distance under the two 
cosmologies at $z=4.35$ is 1.38}&4.35&$-$28.7&2.5&10.2& 230& 21&12.4&1.5&2.0& 1950&1.1\\  
\hline
\end{tabular}
\end{minipage}
\end{table*}\label{table_inferred_properties}

\subsection{Inferred astrophysical properties}\label{subsection_astrophy_props}
Shown in Table 2 are physical properties of the detected sources
derived from the optical luminosities and submillimetre fluxes reported here.
We have assumed that the measured submillimetre flux arises  
from thermal emission from warm dust, and that the dust is heated solely 
by the UV flux from massive star-formation.  
The former can be justified, at least in part, by the radio-selection criterion
that we used to define the parent sample. In addition, we will 
be able to combine the $850\mu m$ fluxes measured here with $1.25$mm 
fluxes that have been measured in a parallel $1.25$mm MAMBO survey 
\cite{Omont01}  running at the IRAM-30m 
(Priddey {\it et al.}, in prep.).
The latter assumption is more difficult to justify, 
as it is possible that a fraction of the rest-frame UV absorbed by 
the dust comes from the central AGN. The size of the AGN contribution is 
difficult to estimate as it depends on 
the source geometry (eg. Sanders {\it et al.}\shortcite{Sanders89}), 
about which we have little knowledge. 
Observations of local, ultra-luminous infrared 
galaxies (ULIRGs, with 
$L_{FIR}> 10^{12}L_{\odot}$), arguably the most likely low-redshift 
counterparts of the quasar host galaxies, have been interpreted as being
suggestive that between 70-80\% of ULIRGS are powered predominantly 
by star-burst activity, with at least 
half the sources observed probably have simultaneous star-burst and AGN
activity on the physical size scales of a few kpcs 
(eg. Genzel {\it et al.}, \shortcite{Genzel98}).
An additional unknown at this point is the extent of gravitational lensing 
and thus source magnification. 
The incidence of gravitational lensing amongst high luminosity
quasars is around 1\% for samples of bright quasars
\cite{Kochanek93}. Studies of previous samples have, however, a
median redshift of only $z\sim2$. At $z\sim4$ the probability of a 
quasar being lensed by a foreground galaxy is a factor of $\sim2$
higher due to the larger optical depth to lensing
\cite{Turner84,Fukugita92}. This qualitative argument is in rough 
agreement with the empirical result that only one of the 49 $z>4$ APM
BRI quasars, the $z=4.43$ APM quasar BRI B0952$-$0115 with
$S_{850\mu m} = 14$mJy (Buffey {\it et al.}, in prep) is known to
be gravitationally lensed.   It therefore seems unlikely that a
large number of the submillimetre-bright source detections reported
here are lensed. High resolution optical imaging of the whole sample
would be needed to confirm this.   
At present we have no evidence to support or refute the possibility of 
magnification, and thus assume that there is none.

\subsubsection
{Astrophysical properties inferred from the optical AGN luminosity} 
An advantage of targeting quasars for a study
of high-redshift (sub)millimetre sources is that we are able to use 
the optical luminosity of the AGN to constrain the mass of the 
black hole powering the AGN, from which we can estimate 
the mass of the surrounding host galaxy.
If we assume a bolometric correction from the $B$ band of 
$L_{\rm bol}/L_B=12$ \cite{Elvis94}, and that 
the black hole is accreting at the Eddington rate, 
then there exists a simple relation between black hole mass, $M_B$, 
and its bolometric luminosity, $L_{\rm bol}$:
\begin{equation}
L_{\rm bol}=\frac{4\pi G m_{\rm p} c M_{\rm bh}}{\sigma_{\rm T}}
=3\times10^{13}\frac{M_{\rm BH}}{10^9\Msun}\Lsun,
\label{eqn_eddlum}
\end{equation}
where $\sigma_T$ is the Thomson scattering cross-section and 
$m_{\rm p}$ the proton mass.

The accretion rate can be written in terms of the luminosity and
an efficiency parameter $\epsilon$, for which a conservative estimate
is $\epsilon_{0.1}\equiv\frac{\epsilon}{0.1}\approx1$
(\cite{Rees84}):

\begin{equation}
\dot{M}_{\rm acc}=\frac{L}{\epsilon c^2}
=\frac{22}{\epsilon_{0.1}}\frac{M_{\rm BH}}{10^9\Msun}
\Msunpyr. 
\label{eqn_accrate}
\end{equation}
While Equation \ref{eqn_accrate} holds, the growth of black hole mass
and luminosity is exponential, with an $e$-folding time-scale 
$\tau_{\rm acc}=M_{\rm BH}/\dot{M}_{\rm acc}=
\epsilon_{0.1}\times4.5\times10^7\yr$.
Studies in the local Universe \cite{Magorrian98,Gebhardt00}
have shown that there exists a linear relation between the mass of 
non-active black holes and the stellar mass of the surrounding galactic 
bulge at the current epoch. If we assume that there is some imprint
of this relationship at early epochs 
(however, see Omont {\it et al.} \shortcite{Omont01} and Priddey {\it
et al.} (in prep)),  
then we can obtain a measure of the mass of the final (ie. $z=0$) quasar host 
galaxy (Column 7 of Table 3) using a value of 
$M_{\rm BH}\approx0.005M_{\rm sph}$ from Gebhardt {\it et al.} 
\shortcite{Gebhardt00}.
It is clearly possible that at $z=4$, the ratio of the black
hole to spheroid mass is different from the local value.
For example Kauffmann \& Haehnelt (2000) predict that the host masses will
less massive by a factor of between 5-10 at $z=2$ as compared to $z=0$.
Work by Rix {\it et al} (1999) on the magnitudes of
lensed host galaxies of lensed
quasars also suggests that the masses of quasar hosts are
low at high redshift however although this is based on the observed
rest-frame B band magnitudes, these observations could be affected 
by dust obscuration, and hence the inferred masses may be underestimated. 
In the long term, one would hope that it will be possible to 
determine the dynamical masses
of quasar hosts at high redshifts so that the
issue can be studied more directly.

These estimates represent lower limits to the mass of
the central engine, as we observe the quasars
while they are still accreting. It seems unlikely for a number of reasons, 
however, that the masses grow much larger than this.
Firstly, the inferred masses are large, comparable to the most 
massive non-active 
black holes ($\la10^{9.5}\Msun$) at the cores of nearby galactic bulges 
\cite{Gebhardt00}. 
Secondly, the accretion rates required to sustain these 
luminosities (Table 3, Column 6)
are extreme--- and encroaching upon ($\sim10$\% of) 
the star-formation rates inferred from
the far-infrared luminosity (Column 11). 
Thus, the gas supplies in the galaxy are being rapidly diminished, at a 
rate too high to sustain either processes for much longer than $1\Gyr$ 
-- indeed 
Priddey \& McMahon determine a median gas consumption timescale of 
$0.1\Gyr$ \cite{Priddey01}.
Although we have no knowledge of the precise accretion history of the 
AGN, we can derive a time taken to assemble the inferred black masses
of around $0.5--1\Gyr$ if we assume that the accretion rate is 
Eddington-limited throughout the growth of the black hole
and that the seeds from which the black holes form are  
$M\sim10^{3-6}\Msun$ (Haehnelt, Natarajan \& Rees, 1998).

\subsubsection{Astrophysical properties inferred from the submillimetre
fluxes} 
Turning to the host galaxy properties derived from    
submillimetre flux observed at frequency  $\nu_o=\nu_{rest}/(1+z)$, 
the dust mass given in Column 9 of Table 3
is determined using
\begin{equation}
M_D=\frac{S(\nu_o)D^2_L}{(1+z)\kappa_D(\nu_r)B_{\nu}(\nu_r,T_D)}.
\end{equation}
where we assume that the emissivity $\kappa$ is a power-law function of
frequency, and we adopt the normalisation of Hildebrand (1983):

\begin{equation}
\kappa_D(\nu)=\kappa_0\left( \frac{\nu}{2.4 {\thinspace\rm THz}}\right) ^{\beta},\label{eqn_kappa}
\end{equation}
where $\kappa_0=18.75\cm^2\g^{-1}$ is the value determined at 
$\lambda=125\mu$m. For the dust temperature and emissivity index, we use
$T_D=40$K and $\beta=2.0$, derived by Priddey \& McMahon \shortcite{Priddey01}
from a composite submillimetre spectrum of z$\ga$4 quasars.
The far-infrared luminosity is calculated by integrating under the 
thermal, grey-body spectral energy distribution, 
\begin{equation}
L_{\rm FIR} = 2\pi h c^2 {\left( \frac{k T}{h c}\right)}^{4+\beta}
\Gamma(4+\beta)\zeta(4+\beta),
\end{equation}
where again we assume $T=40$K, $\beta=2.0$, $\Gamma$ is the
Gamma-function and $\zeta$ is the Riemann-Zeta function.  
If we assume that all the energy re-radiated by the dust originates 
from stars, then the far-infrared luminosity can be used to 
estimate the current star-formation rate:

\begin{equation}
\dot{M_*}=\Psi\times\frac{L_{\rm FIR}}{10^{10}\Lsun}\Msunpyr.
\label{eqn_sfr}
\end{equation}
The exact value of $\Psi$  depends sensitively upon the 
star-formation history of the host galaxy, the stellar 
Initial Mass Function (IMF) and the fraction of starlight absorbed 
by the dust cloud. We adopt the simplifying assumptions that the starformation 
rate is constant and that the cloud-covering factor is unity. Then, 
for a Salpeter IMF with upper and lower mass cut-offs of 
$100$\Msun and $0.3$\Msun respectively, $\Psi\approx1$ and is approximately
constant for ages $\ga10^8$yr.  
The total time required to form the final stellar mass (Column 7) 
inferred from the black hole mass is given in Column 12 of 
Table 3, and is roughly $1\Gyr$.
The starformation history derived from the chemical/dynamical model of 
elliptical galaxy evolution developed 
by Fria\c{c}a \& Terlevich \shortcite{Friaca98} has just such a form--- 
a constant star formation rate lasting for around $1\Gyr$. 
Observational support for long
periods of sustained high starformation rates 
is reported in Abraham {\it et al.} 
\shortcite{Abraham99} (see Figure 6). 
For an alternative interpretation, 
however, see Omont {\it et al.} \shortcite{Omont01} who argue for shorter 
burst timescales.

We have two time-scales describing the fate of gas in the
evolving host galaxy: a time-scale for the conversion of gas into stars, 
$\sim1\Gyr$, or gas accretion onto a super-massive central black hole 
with an accretion time of the same order. 
This may be suggestive of a possible synchrony between formation of 
the AGN and its host galaxy, however we 
leave a more detailed and speculative discussion of the 
inferred quasar and host galaxy properties to a future paper. 

Throughout, we have assumed that the dominant heating source is 
starformation in the quasar host galaxy. At present, however, it is not 
possible to determine conclusively whether this is indeed the case. 
If instead, we were to that assume the AGN contribution to the dust 
heating cannot be neglected, then (a) for a fixed dust temperature the
inferred starformation rate would fall, with a corresponding increase in the 
formation time for the massive spheroid (b) a hotter dust component heated 
directly by the AGN would contribute significantly to the overall
far-infrared luminosity, and thus inaddition, smaller dust masses as well 
as lower starformation rates and longer timescales would be inferred.
Observations of samples, such as this SBQS sample, at shorter wavelengths
using SIRTF will reveal any warmer dust component.

\begin{figure}
\centering
\vspace*{6.5cm}
\leavevmode
\includegraphics{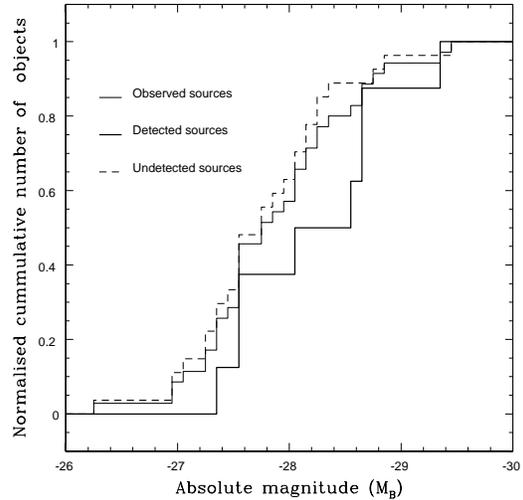}
\caption{Cumulative distribution of the fraction of 
detections (heavy solid line), non-detections (dashed line) and observations
(light solid line) as a function of absolute magnitude. 
There is some indication of a difference in the cumulative 
distribution, however using the K-S test, it can be shown that the 
null hypothesis that the magnitudes of the detected sources and the 
non-detected sources are similarly distributed cannot be rejected at 
the 90\% level.}\label{fig_kstest_mB}
\end{figure}

\begin{figure}
\centering
\vspace*{6.5cm}
\leavevmode
\includegraphics{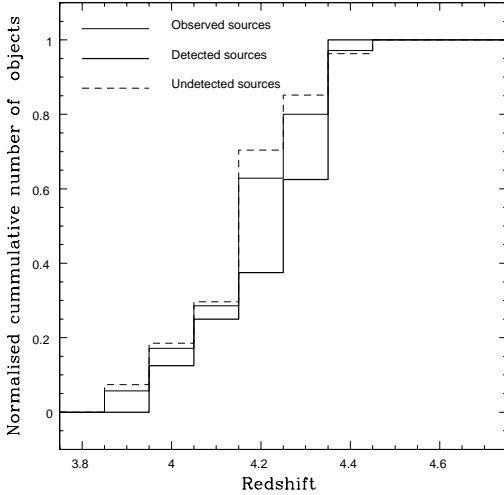}
\caption{Cumulative distribution of the fraction of 
detections (heavy solid line), non-detections (dashed line) and observations
(light solid line) as a function of optical redshift.
Again, using the K-S test, we cannot reject the null hypothesis that the 
redshift distributions of the detected and non-detected sources are
the same.}\label{fig_kstest_z}
\end{figure}

\subsection{Optical/submillimetre correlations}\label{section_optsub_correls}
\begin{figure}
\centering
\vspace*{6.5cm}
\leavevmode
\includegraphics{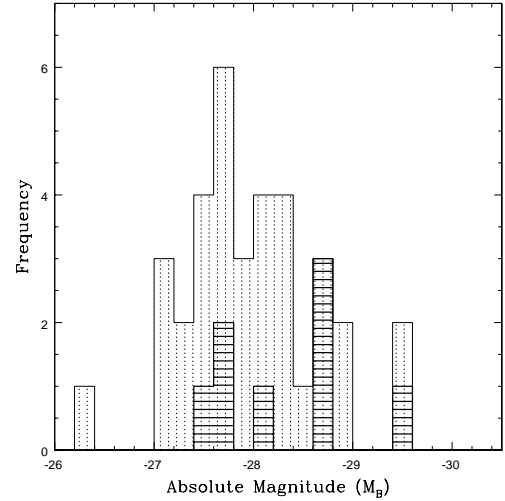}
\caption{Histograms of the absolute magnitudes ($M_B$) of the 
observed sample (dotted shading) and the detected sample (solid 
horizontal shading).}\label{fig_hist_mB}
\end{figure}

\begin{figure}
\centering
\vspace*{6.5cm}
\leavevmode
\includegraphics{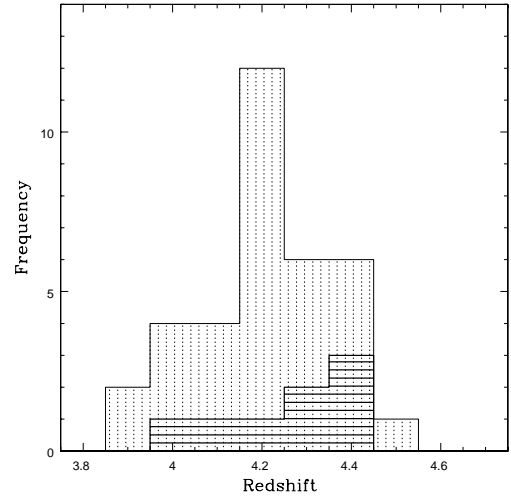}
\caption{Histograms of the redshift distribution of the observed (dotted 
shading) and the detected sample (solid horizontal shading).}\label{fig_hist_z}
\end{figure}

With a sample of \obs observed sources, it becomes 
meaningful to start to make both qualitative and quantitative assessments 
of the trends in the detection rates with source 
redshift and magnitude. Shown in Figures 
\ref{fig_kstest_mB} and \ref{fig_kstest_z} are the cumulative 
magnitude and redshift distributions for the observed, detected 
and non-detected subsamples. To assess quantitatively the degree of similarity 
between the distributions we use the Kolmogorov-Smirnoff (K-S) test (eg. 
Barlow \shortcite{Barlow}). In Figure \ref{fig_kstest_mB} we note that 
there is some indication of a difference between the detected and 
non-detected cumulative magnitude distributions however, formally, 
the difference is not statistically significant, and so we cannot 
reject the null hypothesis that the two samples are drawn from the 
same population. The magnitude and redshift distribution of the two 
samples are shown in a slightly different way in Figures \ref{fig_hist_mB} and
\ref{fig_hist_z}. 

\begin{figure}
\centering
\vspace*{6.5cm}
\leavevmode
\includegraphics{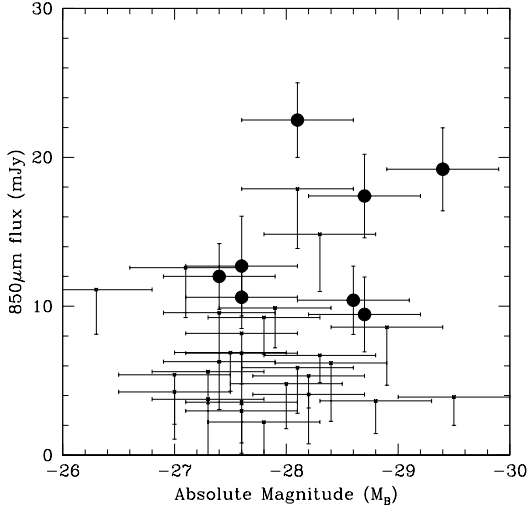}
\caption{A plot of $850\mu m$ flux vs. $M_B$. Filled circles denote
source detections reported in this paper, while crosses denote upper limits to 
non-detections (where plotted flux = ``signal'' + $1.645\sigma$ (90\% confidence)). 
The flux error bars denote $1\sigma$ errors, as given in Table 2, 
whilst magnitude error bars, $\sigma_{MB}\pm 0.5$, are a measure of the 
combined systematic and random error in the measurement and derivation of 
the absolute magnitudes. \label{fig_flux_mB}}
\end{figure}

Plotted in Figure \ref{fig_flux_mB} are the $850\mu m$ fluxes and 
absolute magnitudes for the detections presented in this 
paper, as well as upper limits 
($1.65\sigma$ (90\% confidence level) + ``on-source signal'') for 
non-detections. Error bars for both detections and non detections are
$1\sigma$.  
It is difficult to make a quantitative assessment of 
the significance of the optical/submillimetre magnitude correlation, 
given the significant fraction of upper limits in the sample.   
There are two competing physical processes that would produce a 
statistically significant correlation between detected submillimetre flux 
and absolute quasar luminosity. Firstly, the most luminous quasars reside in 
the most massive host galaxies, in which there is likely to be ongoing 
star-formation -- the larger the host galaxy, the greater the potential for 
massive bursts of star-formation, and thus more dust and more UV photon flux
with which to heat the dust. More quantitatively, 
using Equations \ref{eqn_eddlum}, \ref{eqn_accrate} and \ref{eqn_sfr}, 
\begin{equation}
\frac{L_{\rm FIR}}{\nu L_B}\approx\frac{0.1}{\Psi\epsilon_{0.1}}
\frac{\dot{M_*}}{\dot{M_{\rm acc}}}.
\end{equation}
If we observe these objects just before the end of their
accretion lifetime, when the accretion rate encroaches upon the star
formation rate, $\dot{M_{\rm acc}}\sim\dot{M_*}\Rightarrow 
L_{\rm FIR}\sim \nu L_B$ \cite{Priddey01}. 
Secondly, the more luminous the 
central quasar, the greater its UV-output, and thus 
the larger the power that may be absorbed by dust, and re-emitted in the 
rest-frame far-infrared. 

A plot of the source magnitude vs. redshift for detections 
and non-detections also fails to reveal a preferential range of 
redshifts/magnitudes for which the number of detections is 
high. It must be 
cautioned, however, that the number of sources in any one bin is small, 
so the overall statistical significance of this result is low.


\begin{figure}
\centering
\vspace*{6.5cm}
\leavevmode
\includegraphics{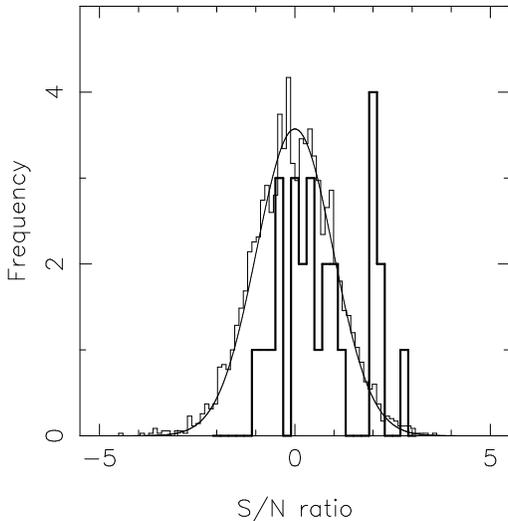}
\caption{Histogram of the S/N ratios of all non-detections in the current
sample: heavy-lined histogram denotes S/N ratios evaluated for the
on-source pixel. Superposed (light-line) is the outline of the 
S/N distribution for all functional off-source pixels, over which is plotted a
Gaussian approximation to the off-source distribution. 
Note that the mean of the off-source distribution is zero, and the 
distribution itself very nearly Gaussian. This suggests that the    
sky-offsets are being removed correctly.\label{fig_sn_hist}}
\end{figure}

In addition, a potential systematic bias arises
from the fact that the absolute $B$-band magnitudes 
have been derived from the measured $R$-band APM magnitudes. 
As discussed in Section 3.1, calculating an optical luminosity from
an $R$ magnitude is complicated for sources at $z\ga4$ because the
band is contaminated by redshifted Lyman-$\alpha$ emission, and at
higher redshifts, intergalactic absorption. 
We introduced a correction to account these effects, and found that
it worked well on average, but that the scatter about the average
was $\sigma_M\sim0.5$. 
Obtaining optical spectra and/or $K$-band photometry for all the 
targets may ultimately help to alleviate this.

\subsection{Non-detections -- an assessment of the contribution of the
$z\sim 4$ population to the submillimetre-wave background}
Shown in Figure \ref{fig_sn_hist} is the signal-to-noise
distribution of the non-detections (solid line) -- this has a clear 
non-zero mean. More quantitatively, the weighted mean of 
fluxes of the observed non-detections is $2.0 \pm 0.6$mJy, a quantity 
that we can evaluate because our survey has a sensitivity that is 
consistent between sources to within a factor of $\sim2$. 
The mean can therefore be used to provide a statistical measure of the 
submillimetre flux of the underlying radio-quiet quasar host 
galaxy population. The derived value is consistent with the earlier 
estimate of $\sim3$mJy based on the deeper observations of a sample 
6 z$\ga$4 radio-quiet quasars \cite{McMahon99}, and is 
also around the value that one might expect for an Arp220-like
ultraluminous infrared galaxy at $z\sim$4.5
(Figure \ref{fig_detectability} -- 
the value obtained is between the curves predicted by the two different
cosmological models). If we assume a source density for $z>4$ quasars of 
1 per $500$deg$^{-2}$, and a mean flux as determined above, then 
the contribution of the quasar host galaxies to the
 $850\mu m$ background is negligible. 

\section{Followup work}
Whilst useful for identifying the most submillimetre-bright sources, the 
detection of objects at a single wavelength clearly provides only 
limited insight into the detailed physical properties of 
the {\it individual} sources. 
Given the strong dependence of the inferred far-infrared luminosity on
dust temperature, and that star-formation rates are derived 
from the far-infrared luminosity,  it is important to 
determine the far-infrared spectral energy distributions 
for individual objects if we are to obtain independent measures of the 
star-formation properties and potential of the quasar host galaxies. 
We have started followup programs at several wavelengths: 
(a) a parallel program at $1.25mm$ using the IRAM-30m which, when combined 
with (b) $450\mu m$ measurements using SCUBA at the JCMT  will be used to 
constrain the spectral energy distributions and thus dust temperatures
and far-infrared luminosities (c) CO observations using
millimetre interferometers to measure the gas mass, simultaneously  
setting upper limits to the physical extents of the high-z host galaxies,
are also in progress along with (d) VLA observations to search for 
the signatures of starformation or weak radio-sources \cite{Carilli01b}.  

Our approach to targeting starformation at high-redshift 
has been to preselect candidates that 
are likely to have massive molecular gas reservoirs, as inferred from 
the presence of dust. 
At some redshift, however, it may become more efficient to 
search directly for the gas reservoir itself, as the presence of
significant quantities of dust preselects for objects in which at 
least one massive starburst of star-formation must already have taken place. 
 
A second selection effect arises from the original definition 
of our sample. 
By basing our sample on the most optically-bright quasars, we tacitly 
assume that the quasar and the host galaxy are coeval. Using 
arguments based on timescales given in Section \ref{subsection_astrophy_props} 
above, we have shown that this may be the case. The evidence, however, is not 
conclusive, and indeed depends on a number of assumptions, not least of 
which being that the rest-frame far-infrared luminosity 
is powered solely by star-formation, and that there is no 
gravitational lensing of the quasar and its host galaxy.   
With the combination of measurements of the CO, dust and optical 
properties, the physical parameters that we can derive from these 
measurements and a sample of statistically significant size, we 
will be able to investigate in a more quantitative manner.

We are currently undertaking a parallel search for 
submillimetre-bright quasar host galaxies at redshifts $z\sim$ 2. 
The elapsed time interval between $z\sim4$ and $z\sim2$ 
is around $1.5\Gyr$, corresponding to a doubling of the age of the 
Universe at $z\sim4$. Given the gas masses
observed in these radio-quiet quasar host galaxies, and 
the high star-formation rates that
we infer from the FIR luminosity (Table 3), 
one might expect that  
if the formation redshift of quasar host galaxies is high,
a sizeable fraction of any molecular gas reservoir 
would be used up between $z\ga4$ and 
by $z\sim2$. The submillimetre-wave
properties of the two samples should therefore be quite different.
We note, however, that the e-folding lifetime of the quasar ($50\rm{Myr}$) is
considerably shorter than this interval (again, see Table 3), 
and so the quasar population that
we see at $z\sim2$ may not be the same as the $z\sim4$ population. 
Any global evolution in submillimetre-wave luminosity 
between $z\ga4$ and $z\sim2$ host galaxies could in this
case reflect variation in the underlying cosmological processes, such as 
merger rates, dynamical timescale and availability of cold gas,
driving the evolution of the population \cite{Kauffmann00}.

\section{Summary}
We have observed a total of 38 z$\ga$4 radio-quiet quasars at 
$850\mu m$ using the JCMT. We have identified a total of 8 new 
submillimetre-bright sources with $S_{850\mu m} > 10$ mJy including  
a previously ambiguous detection made at the IRAM-30m at 1.25 mm. 
The observations were carried out under modest observing conditions
(sky transmission $\sim65$\%), but even so the integration time per
target was around 0.5 hour, giving an overall detection efficiency
of 0.5 sources per hour of integration.
This method, observing preselected targets in photometry mode,
thus compares favourably with the complementary strategy of
blank-field mapping. 
These observations triple the number of known z$\ga$4 objects for which 
CO observations should be possible in a finite amount of observing time
to $N>10$. As such, we now have a submillimetre-bright sample of 
statistically significant size, 
for which we possess accurate identifications and spectroscopic
redshifts, for which we can now start to study 
the star-formation properties in detail.  
We analysed a new optical spectrum of the BAL PSS J1048$+$4407, one of the 
sources detected, and find that the redshift is quite different from that
used for CO observations of Guilloteau et al.,\shortcite{Guilloteau99}, 
which may explain the observed lack of CO from this object. 
The strong 850$\mu m$ emission in contrast to the lack of 
1.35mm emission detected in PdB observations 
suggests strongly that the much higher angular resolution obtained with the 
interferometer is causing significant fraction of the emission to be resolved
out -- the (sub)millimetre- source is extended.  

Of the remaining 30 objects observed, 27 were observed down to a 
sensitivity of $3\sigma < 12$mJy and not detected. Based on this 
sample, we find that our detection rate is comparable to that measured
by Omont {\it et al.},\shortcite{Omont96b} 
at 1.25 mm using 30-m, scaled to match the appropriate
flux limits. At present, the sample (detections and non-detections) is 
too small to determine formally whether there is a range of 
absolute magnitudes and redshifts for which the rate of detection is high, 
however there may be some indication that the magnitude distribution of the
detections and non-detections differ. We determined a
fiducial flux of $2.0\pm 0.6$mJy 
for the non-detections, which we believe to be a statistical measure of the 
submillimetre flux of the more submillimetre-faint quasar host population.

\section*{Acknowledgments}
We thank the support scientists and telescope support specialists 
at the JCMT, in particular Gerald Moriarty-Schieven and 
Iain Coulson, the ``displaced observer'' for the care and 
attention afforded to the observations and Tim Jenness for very useful 
discussions on SURF and ORACDR. KGI, RSP, CP and RS acknowledge a 
PPARC fellowship and PPARC studentships respectively, and RGM acknowledges
the support of the Royal Society.
This work was partially funded by the EC TMR FMRX-CT96-0068 and EARA.
The JCMT is operated by JAC, Hilo, on behalf of the parent organisations 
of the Particle Physics and Astronomy Research Council in the UK, the 
National Research Council in Canada and the Scientific Research Organisation 
of the Netherlands.

\bibliography{highz_astronomy}
\bibliographystyle{unsrtnat}

\label{lastpage}

\end{document}